\documentclass[article,twocolumn,preprintnumbers,amsmath,amssymb,floatfix,superscriptaddress]{revtex4}
\usepackage{graphicx}
\usepackage{rotating}
\usepackage{comment}
\usepackage{dcolumn}% Align table columns on decimal point
\usepackage{makeidx}
\usepackage{color}
\usepackage{amsmath}
\usepackage{amsfonts}
\usepackage{wrapfig}
\usepackage{braket}
\usepackage{bm}
\usepackage{float}
\usepackage{hyperref}
\hypersetup{colorlinks=true, linkcolor=blue, citecolor=blue, urlcolor=blue}

\usepackage[normalem]{ulem}
\usepackage[usenames, dvipsnames]{xcolor}

\begin{document}

\title{
Spontaneous skyrmionic lattice from anisotropic symmetric exchange \\ in a Ni-halide monolayer 
}

\author{Danila Amoroso}
\affiliation{\footnotesize Consiglio Nazionale delle Ricerche CNR-SPIN, c/o Universit\`a degli Studi “G. D’Annunzio”, I-66100 Chieti, Italy}
\author{Paolo Barone}
\affiliation{\footnotesize Consiglio Nazionale delle Ricerche CNR-SPIN, c/o Universit\`a degli Studi “G. D’Annunzio”, I-66100 Chieti, Italy}
\author{Silvia Picozzi}
\affiliation{\footnotesize Consiglio Nazionale delle Ricerche CNR-SPIN, c/o Universit\`a degli Studi “G. D’Annunzio”, I-66100 Chieti, Italy}

%\date{\today}
%\newcommand{\}{}

%\pacs{Valid PACS appear here}

\begin{abstract}
{\bf 
Topological spin structures, such as magnetic skyrmions, hold great promises for data storage applications, thanks to their inherent stability. In most cases, skyrmions are stabilized by magnetic fields in non-centrosymmetric  systems  displaying  the  chiral  Dzyaloshinskii-Moriya exchange interaction, while spontaneous skyrmion lattices have been reported in centrosymmetric itinerant magnets with long-range interactions. Here, a spontaneous anti-biskyrmion lattice with unique topology and chirality is  predicted  in  the monolayer of a semiconducting and centrosymmetric metal halide,  NiI$_2$. Our first-principles and Monte Carlo simulations reveal that the anisotropies of the short-range symmetric exchange, when combined with magnetic frustration, can lead to an emergent chiral interaction that is responsible for the predicted topological spin structures. The proposed mechanism finds a prototypical manifestation in two-dimensional magnets, thus broadening the class of materials that can host spontaneous skyrmionic states.
}
\end{abstract}

\maketitle

Magnetic skyrmions are localized topological spin structures characterized by spins wrapping a unit sphere and carrying an integer topological charge $Q$~\cite{tokura_review,fert_review}. Their topological properties ensure the inherent stability that makes them technologically appealing for future memory devices~\cite{parkin_racetrack}. Non-coplanarity in the direction of the 
spin magnetic moments at different lattice sites is a necessary - albeit not sufficient - ingredient to obtain a net scalar spin chirality $\bm{S}_i \cdot({\bm S}_j\times \bm{S}_k)$, in turn related to the topological invariants that characterize most of the appealing properties of skyrmions~\cite{tokura_review,fert_review}. While the interplay of competing magnetic interactions may often lead to non-coplanarity, localized skyrmion-like magnetic textures with fixed chirality are generally believed to arise from the Dzyaloshinskii-Moriya (DM) interaction, driven by spin-orbit coupling (SOC) in systems lacking space-inversion symmetry. Such short-range antisymmetric exchange  interaction, in fact, fixes one specific rotational sense of spins, thus imposing a well-defined chirality to non-collinear and non-coplanar spin textures ~\cite{robler_2006, nagaosa_2009, stefan_2011, tokura_report_2012, kristian_prb_2014, batista_2015, tokura_nanotech_2020}. Complex helical orderings, including skyrmion-lattice states~\cite{okubo_prl,maxim_2015,landau_2016,batista_2016,tokura_2019}, can also arise from competing exchange interactions on geometrically frustrated centrosymmetric lattices, such as triangular or Kagome ones. These states, that are metastable in classical isotropic systems, can be stabilized by applied fields, easy-axis magnetic anisotropy or long-range %\u2013 
dipole-dipole
and/or Ruderman-Kittel-Kasuya-Yosida (RKKY) %\u2013 
interactions and thermal or quantum fluctuations
~\cite{okubo_prl,maxim_2015,landau_2016,batista_2016,tokura_2019}. However skyrmionic spin structures can here manifest with various topologies, as there is no mechanism determining \emph{a priori} their topology and chirality, as opposed to non-centrosymmetric chiral magnets with DM interaction ~\cite{tokura_current_2014,adv_mater_2Sk, current_induced_2017,capic_2019,ingrid_2019}. Furthermore, while skyrmion states are usually stabilized by external fields in both centrosymmetric and non-centrosymmetric materials, a spontaneous skyrmion-lattice ground-state has been proposed so far only in itinerant magnets displaying amplitude variations of the magnetization\cite{robler_2006}, where its microscopic origin was attributed to long-range effective four- and higher-spin interactions that arise from conduction electrons ~\cite{stefan_2011, motome_1, motome_2}.\\

{\bf Results}\\
{\bf Skyrmionic lattice in NiI$_2$.} We report, here, a spontaneous high-$Q$ anti-skyrmion lattice and a field-induced topological transition to a standard skyrmion lattice in a novel 2D-magnet, the centrosymmetric NiI$_2$ monolayer.
By performing Density Functional Theory (DFT) and Monte-Carlo (MC) simulations, %See Supplementaryfor technical details. 
we show that the anisotropic part of the short-range symmetric exchange can fix a unique topology and chirality of the spin texture, 
thus leading to the spontaneous formation of thermodynamically-stable skyrmionic lattice, in absence of DM and Zeeman interactions, but assisted only by the exchange frustration. 

\begin{figure*}[t]
\centering
\includegraphics[width=\textwidth]{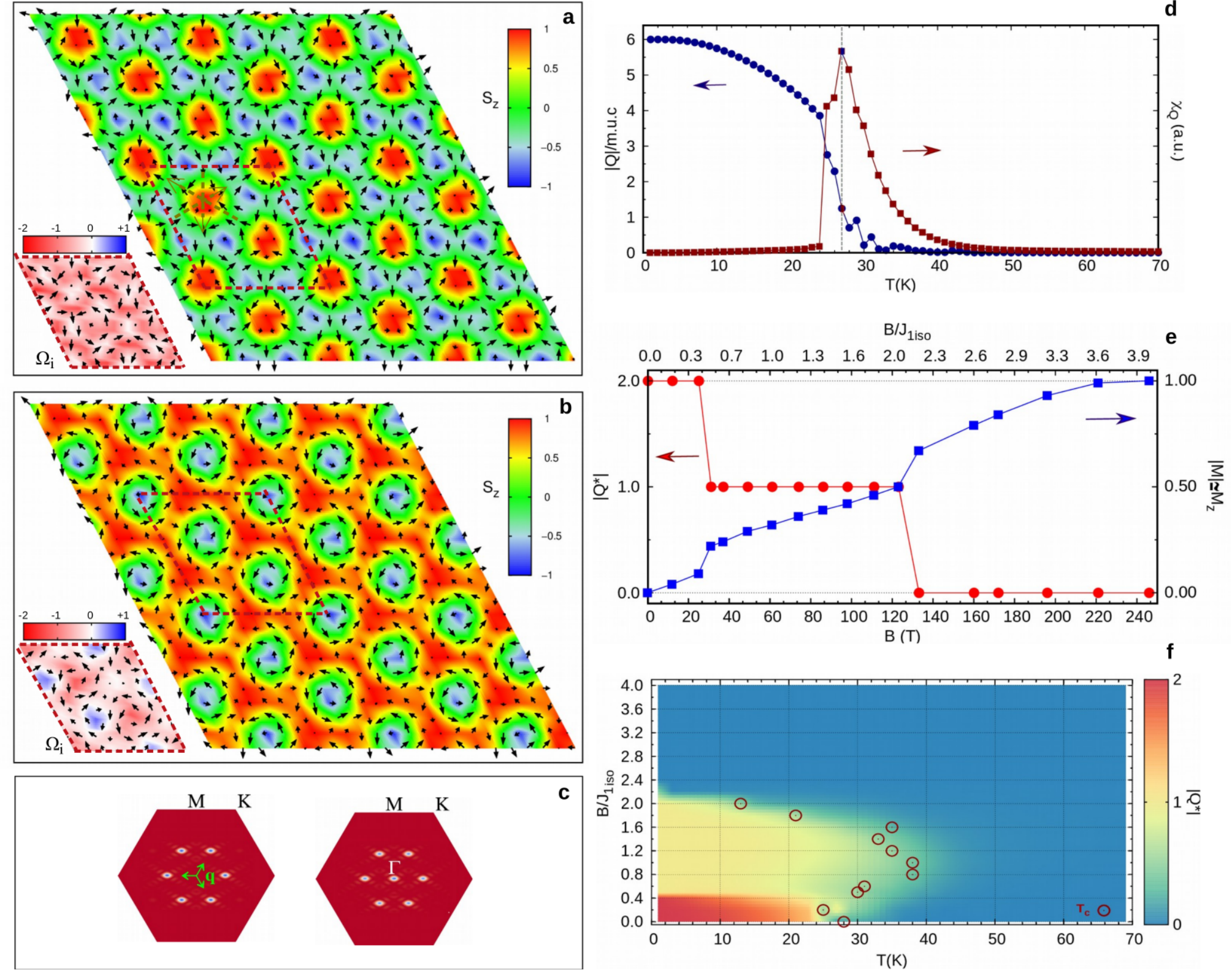}
\caption{\footnotesize  {\bf Spin structures and field-induced topological  transition of the thermodinamically phases in monolayer NiI$_2$.} 
{\bf a, b)} Snapshots of spin configurations 
at $T=1~$K from MC simulations on a 24$\times$24 supercell obtained for $B/J_{1iso}=0$ and $B/J_{1iso}\simeq 1.5$, displaying the anti-biskyrmion and Bloch-skyrmion lattice respectively. Black arrows represent in-plane components of spins, colormap indicates the out-of-plane spin components. The magnetic unit cell is shown with dashed lines; dashed arrows in (a) are guidelines for the eyes to visualize the spins orientation and directions defining the anti-biskyrmion. Insets show correspondent topological charge densities $\Omega_i$ in the selected magnetic unit cell. 
{\bf c)} The spin structure factor $S(\bm q)$, corresponding to a triple-$\bm q$ state with $\bm q_1=(\delta,\delta)$, $\bm q_2=(\delta,-2\delta)$ and $\bm q_3=(-2\delta,\delta)$ in the hexagonal setting, as highlighted by green arrows, with $\delta\simeq 0.125$. $S(\bm q)$ on the left refers to the A2Sk ground-state; $S(\bm q)$ on the right refers to the Sk state with the additional peak at $\Gamma$ reflecting the ferromagnetic component induced by the applied magnetic field. 
{\bf d)} Topological charge $|Q|$ (closed circles) per magnetic unit cell and corresponding topological susceptibility $\chi_Q$ (closed squares) as a function of temperature for $B/J_{1iso}=0$, pointing to a transition temperature T$_c\simeq 28~$K within the used DFT approximations. 
{\bf e)}, Evolution of $|Q|$, normalized to the number of skyrmionic objects in the magnetic unit cell 
	%($3$ both in (a) and (c)), 
and of magnetization $M$ as a function of the magnetic field $B$ at $T= 1~$K. 
{\bf f)}, Phase diagram in the temperature-field plane. Colormap indicates the topological charge densities $|Q|$ as defined in (e). 
	Critical temperatures (T$_c$) for topological phase transitions are also reported in circles.    }
\label{skyrmion_lattice}
\end{figure*}

NiI$_2$ is a centrosymmetric magnetic semiconductor long known for its exotic helimagnetism~\cite{nibr2_1980, nii2_1981, nibr_tokura_2011, nii2_tokura_2013}. It belongs to the
family of transition-metal-based van der Waals  materials %(CrI$_3$ being the prototype) 
recently object of intense research activity 
due to their intriguing low 
dimensional magnetic properties
~\cite{mc_guire, kulish,botana, babu, 2D_hot,CrSe, VSe2, helicity_tune_adv_mat, adv_mater_2d}. A single layer of NiI$_2$
is characterized by a triangular net of magnetic cations and competing ferro (FM)- and antiferro(AFM)- magnetic interactions, 
resulting in strong magnetic frustration (Fig.~S2 in Supplementary). 
Our DFT and MC calculations reveal that NiI$_2$ monolayer displays a spontaneous transition below $T_c\simeq30~K$ to 
a triple-{\bf q} state. This state consists in a triangular lattice of anti-biskyrmions (A2Sk)
characterized by a topological charge $Q=2$ 
with associated vorticity $m=-2$ and helicity $\eta=\pi/2$~\cite{bertrand_2016, kristian_prb_2017, current_induced_2017}, as detailed in Fig.~\ref{skyrmion_lattice}. 
The magnetic cell of such A2Sk lattice
comprises three anti-biskyrmions, each surrounded by six vortices with vanishing net magnetization;
the central spins of the A2Sk have opposite $S_z$ component with respect 
to the vortices centre (in the specific case reported in Fig.~\ref{skyrmion_lattice}(a), spins in the anti-biskyrmion core and in the vortices centers point upward and downward, respectively). 
When a perpendicular magnetic field is applied, 
a conventional Bloch-type skyrmion lattice with $m=1$, $\eta=-\pi/2$, shown in Fig.~\ref{skyrmion_lattice}(b)
is induced for a finite range of applied fields, before a ferromagnetic state is finally stabilized for large fields. Upon the field-induced transition, all spins surrounding one every two downward vortices align to the field, while the in-plane spin pattern remains substantially unchanged. 
A sharp topological phase transition thus occurs as the charge $|Q|$ changes from 2 to 1 at the critical ratio $B/J_{1iso}> 0.4$,
 while it changes to 0 for $B/J_{1iso}\simeq 2.2$, as displayed in Figs.~\ref{skyrmion_lattice}(d,e,f).
The magnetization $\bm M$  also exhibits evidences of these phase transitions, signalled by abrupt changes in correspondence of the critical fields; 
the magnetization saturation, corresponding to all spin aligned with the magnetic field, takes place for $B/J_{1iso} > 3.9$, as shown in Fig.~\ref{skyrmion_lattice}(e). 

A similar sequences of topological phase transitions and the spontaneous onset of the A2Sk lattice
has been previously predicted in frustrated itinerant magnets described by a Kondo-lattice model on a triangular lattice.~\cite{motome_1} 
Nevertheless, the effective interaction between localized spins mediated by conduction electrons, that has been identified as the driving mechanism for the stabilization of the topological spin textures in such itinerant magnets\cite{motome_2, motome_3}, cannot be invoked to explain the A2Sk lattice in semiconducting NiI$_2$. Similarly, the DM interaction is to be excluded, being forbidden by the inversion symmetry of the lattice.
As discussed below, here the A2Sk ground-state, and related field-induced state, directly arise from the anisotropic properties of the \emph{short-range} symmetric exchange. As such, the underlying mechanism is not restricted to itinerant magnets and metallic systems, but rather has a more general validity, as it can apply also to centrosymmetric magnetic semiconductors.\\

{\bf Underlying microscopic mechanisms.}
Magnetic interactions between localized spins $\vec{S}$ can be generally modeled by the classical spin Hamiltonian
\begin{equation}
	H=\frac{1}{2}\sum_{i\neq j}\vec{S_i}{\bm J}_{ij}\vec{S_j}+\sum_i\vec{S_i}{\bm A}_{i}\vec{S_i}
	\label{hamiltonian1}
\end{equation}
where $\bm A_i$ is the on-site or single-ion anisotropy (SIA) tensor, and ${\bm J}_{ij}$ is the exchange tensor. The latter is generally decomposed into 
three contributions: the isotropic coupling term  $J^{iso}_{ij}=\frac13Tr{\bm J}_{ij}$, defining the scalar Heisenberg model 
$H^{iso}=\frac{1}{2}\sum_{i\neq j}J_{ij}\vec{S_i}\cdot\vec{S_j}$; the antisymmetric term ${\bm J}^A_{ij}=\frac12({\bm J}_{ij}-{\bm J^T}_{ij})$, which corresponds to 
the DM interaction and vanishes in the presence of an inversion center on the spin-spin bond (as realized in the systems under investigation); 
the anisotropic symmetric term ${\bm J}^S_{ij}=\frac12({\bm J}_{ij}+{\bm J^T}_{ij})-J^{iso}_{ij}\bm I$, also referred to as the two-site magnetic anisotropy, which is of particular interest here. 
As the DM interaction,  the two-site anisotropy arises from the spin-orbit coupling; while the former favours the canting of spin pairs, the latter tends to orient the spins along given orientations in space. When the principal (anisotropy) axes between different spin pairs are not parallel, an additional frustration in the relative orientation of adjacent spins may lead to non-coplanar magnetic configurations, in analogy with the mechanism devised by Moriya for canted magnetism arising from different single-ion anisotropy axes \cite{moriya1960}.
In Table~\ref{exchange_j} we report such contributions to the exchange coupling for NiI$_2$ monolayer as obtained by DFT-calculations. In order to highlight chemical trends,
we also report results for the isostructural NiCl$_2$ and NiBr$_2$ monolayer systems.
This allows, on the one hand, 
to analyze the effects of different SOC strengths (going from the weak contribution expected in Cl to the strongest one in I) and, on the other hand, to explore the  range of the interactions (going from the localized 3$p$ states in Cl to broader 5$p$ in I). 

\begin{table}[b]
\centering
\begin{tabular}{ccccccc}
\toprule        
	& \multicolumn{2}{c}{$\bm{J_{1iso}}$} & \multicolumn{2}{c}{$\bm {J_{2iso}}$}  & \multicolumn{2}{c}{$\bm {J_{3iso}}$}  \\
	{\bfseries NiCl$_2$} & \multicolumn{2}{c}{-5.1} & \multicolumn{2}{c}{-0.1} & \multicolumn{2}{c}{+1.7} \\
	{\bfseries NiBr$_2$} & \multicolumn{2}{c}{-5.9} & \multicolumn{2}{c}{-0.1} & \multicolumn{2}{c}{+2.9} \\
	{\bfseries NiI$_2$} & \multicolumn{2}{c}{-7.0} &  \multicolumn{2}{c}{-0.3}  & \multicolumn{2}{c}{+5.8} \\
\hline 
\hline 
	& \multicolumn{6}{c}{$ \bm {J^{1}_{2site~anisotropy}}$}  \\
  &  $J_{xx}$  & $J_{yy}$  &  $J_{zz}$  & $J_{yz}$   & $J_{xz}$  &  $J_{xy}$ \\
%\midrule
{\bfseries NiCl$_2$} & 0.0  & 0.0  &  0.0  &  0.0  & 0.0  & 0.0   \\
{\bfseries NiBr$_2$} & -0.1  & +0.1  &  0.0  &  -0.1  & 0.0  & 0.0   \\
	{\bfseries NiI$_2$} & -1.0 & +1.4  &  -0.3  &  -1.4  & 0.0  & 0.0   \\
\hline
\hline
	& \multicolumn{2}{c}{$\bm{\lambda_{\alpha}}$} & \multicolumn{2}{c}{$\bm {\lambda_{\beta}}$}  & \multicolumn{2}{c}{$\bm {\lambda_{\gamma}}$}  \\
        {\bfseries NiCl$_2$} & \multicolumn{2}{c}{-5.1} & \multicolumn{2}{c}{-5.1} & \multicolumn{2}{c}{-5.1} \\
        {\bfseries NiBr$_2$} & \multicolumn{2}{c}{-6.0} & \multicolumn{2}{c}{-6.0} & \multicolumn{2}{c}{-5.7} \\
        {\bfseries NiI$_2$} & \multicolumn{2}{c}{-8.1} &  \multicolumn{2}{c}{-8.0}  & \multicolumn{2}{c}{-4.8} \\
\botrule
\end{tabular}
	\caption{\footnotesize {\bf Exchange coupling parameters for NiCl$_2$, NiBr$_2$ and NiI$_2$ monolayers}.  (upper table) Isotropic first-, second- and third- 
	neigbhor interactions. (middle table) Anisotropic symmetric exchange, or two-site anisotropy, between first nearest-neighbour spins. Values refer to the Ni$_0$-Ni$_1$ pair, with the Ni-Ni bonding vector parallel to the cartesian axis $x$. 
	(bottom table) Principal values (eigenvalues) of the fist-neighbours exchange tensor. 
	Exchange parameters are expressed in term of energy unit (meV). 
	%and are calculated via DFT+U calculations with U=1.8 eV, J=0.8 eV. 
	}
	\label{exchange_j}
\end{table}

As expected, the FM nearest-neighbour exchange interaction becomes larger when moving from Cl to I, as a consequence of the broader ligand $p$ states mediating the superexchange. Interestingly, the resulting AFM third-nearest neighbour interaction is also strongly affected: the 
$J_{3iso}/J_{1iso}$ ratio is $\simeq -0.33, -0.49, -0.83$ for NiCl$_2$, NiBr$_2$ and NiI$_2$, respectively, revealing thus increasing magnetic frustration as a function of the ligand.
A similar trend is observed for the exchange anisotropy, that also increases as the ligand SOC gets stronger along the series.
Indeed, magnetism of NiCl$_2$ results well described by the isotropic Heisenberg model,
as opposed to NiBr$_2$ and NiI$_2$. The strongest effect is predicted for NiI$_2$,
where the two-site anisotropy is one order of magnitude larger than in NiBr$_2$. 
In particular, the $J_{yz}/J_{1iso}$ ratio, measuring the canting of the two-site anisotropy axes from the direction perpendicular to the monolayers, changes from $0.00$ in NiCl$_2$ to  $0.02$ and $0.20$ in NiBr$_2$ and NiI$_2$, respectively. 
This specific anisotropic contribution tends to vanish in the second- and third-nearest neighbour exchange interactions. Moreover, second-nearest-neighbour and beyond third-nearest-neighbour interactions are at least one order of magnitude smaller than $J_1$ and $J_3$; the SIA (whose values are reported in Supplementary-Table~SI)
is also negligible with respect to the two main interactions.
In the following we will therefore focus on the crucial role played by the two-site anisotropy in driving the A2Sk lattice in NiI$_2$.

\begin{figure}[t]
\centering
\includegraphics[width=\columnwidth]{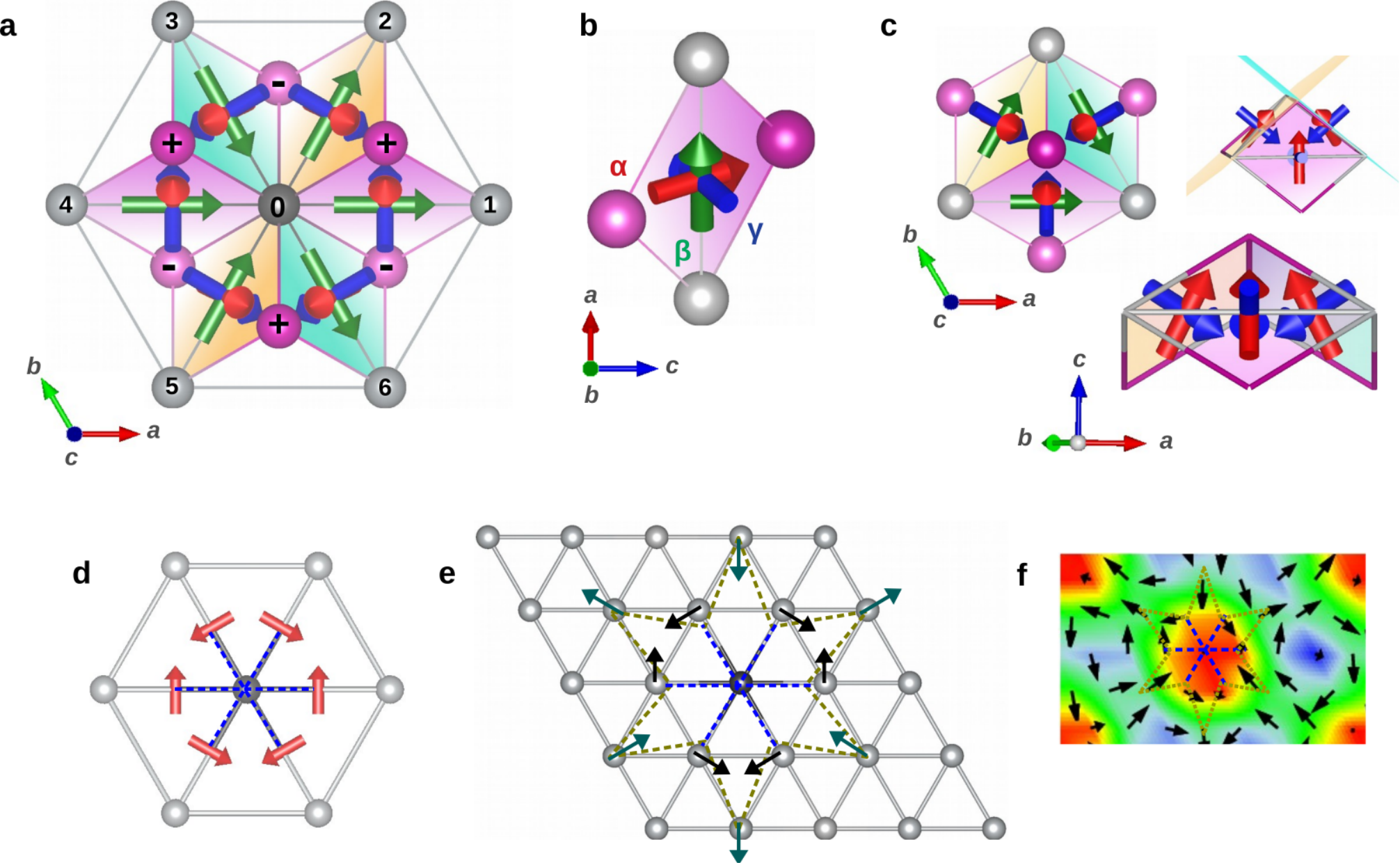}
	\caption{\footnotesize {\bf Schematic visualizations of the magnetic eigenvectors and their relation with the anti-biskyrmion spin structure.} 
	{\bf a)} Top view of NiI$_2$: the six nearest-neighbours Ni surrounding the central Ni atom are labeled with numbers. 
	Plus and minus indicate iodine atoms above and below nickels plane respectively. 
	The $\{\bm\nu_{\alpha}, \bm\nu_{\beta}, \bm\nu_{\gamma}\}$ eigenvector basis set is represented on each Ni-Ni pair.  
	{\bf b)} Lateral view of the Ni-I-Ni-I plaquette and relative orientation of the eigenvectors: $\bm\nu_{\alpha}$ ($\bm \alpha$-red), $\bm\nu_{\beta}$ ($\bm \beta$-green) and $\bm\nu_{\gamma}$ ($\bm \gamma$-blue). 
	{\bf c)} Top and lateral view of the local eigenvectors on the triangular Ni-net to help visualization of the non-coplanarity and non-collinearity in the exchange-tensor principal axes. 
	{\bf d)} In-plane components of the $\bm \nu_\alpha$ eigenvector for each magnetic Ni-Ni pair. 
	{\bf e)} Sketches of the anti-biskyrmion spin structure: spins on the magnetic sites of the nearest-neighbours (black arrows) of the central Ni orient according to 
	the in-plane projection of the non-coplanar principal axes (d);
	spins on the second-nearest neighbour magnetic sites (green arrows) orient following the direction fixed by the interaction accommodating the A2Sk in a spin lattice.
	Dashed gold lines are guidelines for the eyes delimiting the topological spin pattern and 
	dashed blue lines mark the direction of the six nearest-neighbours Ni. 
	{\bf f)} Zoom on the anti-biskyrmion lattice as obtained from the 
	MC simulations: spins orient as drawn in (e).}
\label{structures}
\end{figure}

The exchange tensor discussed until now was expressed in the cartesian $\{x,y,z\}$ basis, where $x$ was chosen to be parallel to the Ni-Ni bonding vector (see Fig~\ref{structures}(a) and Fig.~S1 in Supplementary). 
In order to better understand and visualize the role of the two-site anisotropy, %in the stabilization of the A2Sk lattice, 
it is useful to express the interaction within the 
local principal-axes basis $\{\bm\nu_{\alpha}, \bm\nu_{\beta}, \bm\nu_{\gamma}\}$ 
which diagonalize the exchange tensor, with eigenvalues ($\lambda_{\alpha}, \lambda_{\beta}, \lambda_{\gamma}$), as detailed in Section SI of Supplementary. Principal values of the exchange tensor for the three compounds are reported in Table~\ref{exchange_j}. 
It is also noteworthy that, within this local basis, the exchange interaction could be further decomposed into an isotropic parameter $J'$ and a Kitaev term $K$, 
as in Refs.~\cite{laurent_cri3_2018, laurent_cri3_2020} and Supplementary-Section SI, thus providing another estimate of 
the global anisotropy of the exchange interaction; the $|K/J'|$ ratio in fact evolves as $\simeq 0.00, 0.05, 0.40 $ in NiCl$_2$, NiBr$_2$ and NiI$_2$, respectively.

In Fig.~\ref{structures} we show the principal axes for the most anisotropic system, NiI$_2$: 
$\bm\nu_{\alpha}$ and $\bm\nu_{\beta}$ vectors lie in the 
Ni-I-Ni-I spin-ligand plaquette 
while $\bm\nu_{\gamma}$ is  perpendicular to it (cfr Fig.~\ref{structures}b). 
Therefore, the six $\bm\nu_{\alpha}$ and $\bm\nu_{\gamma}$ vectors, being not parallel to any lattice vector,
introduce a non-coplanar component in the interaction of the spins (Fig.~\ref{structures}c). 
This non-coplanarity is mathematically 
determined by the off-diagonal terms of the two-site anisotropy (see Section SI of Supplementary), arising from the SOC of the heavy I ligand, combined with the non-coplanarity of the spin-ligand plaquettes on each triangular spin-spin plaquette displayed in Fig~\ref{structures}(c).
Such non-coplanarity is the first main ingredient introduced by the two-site anisotropy and necessary to define the net scalar chirality of the spin texture. Its second main effect is also to fix the helicity, driving the in-plane 
orientation of the spins. The topology and chirality of the spin texture is thus well defined. In fact, when looking
at the in-plane projection of the $\bm\nu_{\alpha}$ eigenvector reported in Fig.~\ref{structures}(d), one can observe a direct relation to the anti-biskyrmion spin-pattern represented in Fig.~\ref{structures}(e,f): 
spins on the nearest-neighbour Ni atoms surrounding the central spin orient in plane according to 
the in-plane components of the non-coplanar $\bm\nu_{\alpha}$ vector both in direction and sense [Fig.~\ref{structures}(e,f)]; the accommodation of the anti-biskyrmions in the spin lattice, along with the conservation of a vanishing net magnetization in the system, 
determine then the direction and orientation of the second-neighbor spins, 
and, by consequence, the pattern of the associated magnetic defects (i.e. the vortices) surrounding the A2Sk-core.
In particular, the rotational sense of the spins, i.e. the helicity $\eta$, is determined by the sign of the off-diagonal terms of the two-site anisotropy.
We verified this dependence
by artificially changing the sign of $J_{yz}$ term in the Ni$_0$-Ni$_1$ exchange tensor (that by symmetry relation also affects related terms of the other Ni-Ni coupling, as in Section~SI-Supplementary) in the MC simulation.
The resulting topological lattice still displays 
vorticity $m=-2$, \emph{i.e.} the A2Sk lattice, but with opposite helicity $\eta=-\pi/2$ (see Fig.~S4 in Supplementary).
Moreover, to further verify the unique relation between the two-site anisotropy and the topology of the spin structure, we also 
artificially swapped the exchange interaction of the Ni$_0$-Ni$_2$ and Ni$_0$-Ni$_3$ pairs 
obtaining now a biskyrmion lattice with $m=2$ (see Fig.~S5 in Supplementary). %, i.e. a biskyrmion lattice. 
In this case, an applied magnetic field stabilizes an antiskyrmion lattice ($m=-1$), as shown in Fig.~S5(e,f).
Our results thus demonstrate that the two-site anisotropy 
may behave as an emergent chiral interaction for this class of centrosymmetric systems,
determining a unique topology and chirality of the spin structure. \\

\begin{figure*}[t!]
\centering
\includegraphics[width=\textwidth]{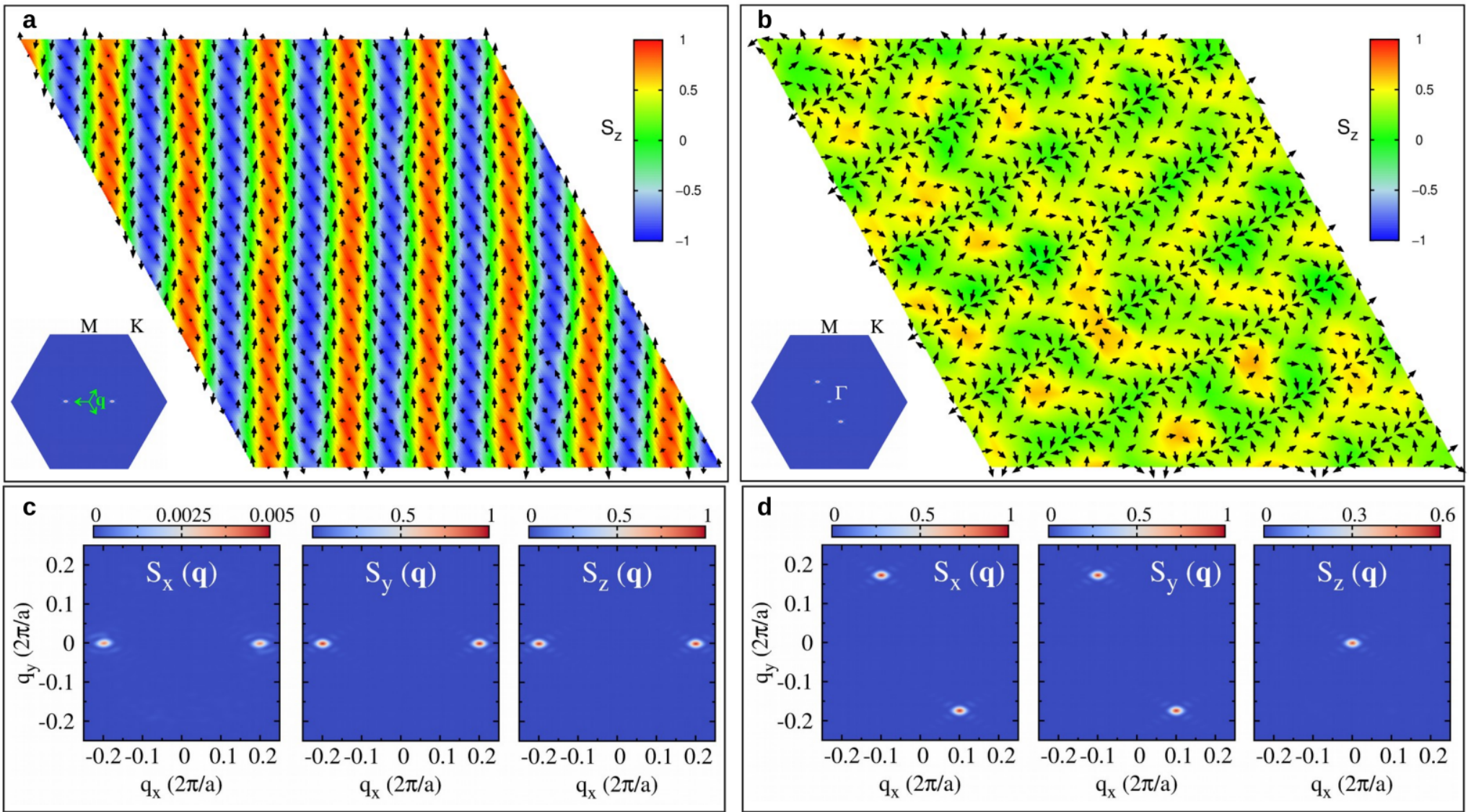}
	\caption{\footnotesize {\bf Magnetic structures for NiBr$_2$ monolayer.} 
	{\bf a,b)} Snapshots of spin configurations at $T=1~$K from MC simulations on a 30$\times$30 supercell obtained for $B/J_{1iso}=0$ and $\simeq 0.3$, respectively. Insets show the spin structure factor S($\bm q$), which corresponds to a single-$\bm q$ state along $\bm q_3 = (-2\delta,\delta)$ in (a) and $\bm q_2=(\delta,-2\delta)$ in (b), with $\delta \simeq 0.10$. 
	{\bf c,d)} Decomposition of the spin structure factor in the cartesian components 
	$S_x(\bm q),~S_y(\bm q)$ and $S_z(\bm q)$. 
		The spin configuration displayed in (a) is composed of a spin-spiral, with spins rotating in a plane perpendicular to the propagation vector $\bm q_3$, and of a spin-cycloidal component, with spins rotating in a plane containing the propagation vector, as highlighted by the small peak in $S_x(\bm q)$. Such non-coplanar helimagnetic single-$\bm q$ state arises from the small anisotropic symmetric exchange reported in Table~\ref{exchange_j}. Under an applied magnetic field, a conical helix state is stabilized (b), where a ferromagnetic component parallel to the field is superimposed to a cycloidal state, as evidenced by in-plane and out-of-plane components of the spin structure factor shown in (d). 
	}
\label{spin_cl_br}
\end{figure*}

{\bf Competing magnetic frustrations.}
It is important to stress that magnetic frustration is a necessary prerequisite for the stabilization of the skyrmionic triple-{\bf q} state, whose topological character is then dictated by the frustrated two-site anisotropy. Indeed, the latter competes with the isotropic exchange that, in the absence of magnetic frustration, would favour collinear spin configurations. This is, for instance, the case of the prototypical 2D ferromagnet CrI$_3$, which displays an easy-axis FM ground-state despite the strong exchange anisotropy~\cite{huang_2017,laurent_cri3_2018}. 
Indeed, Xu and coauthors~\cite{laurent_cri3_2018} 
reported a strong anisotropic symmetric exchange for monolayer CrI$_3$, akin to what we obtained for monolayer NiI$_2$ and consistently with the similar metal-halide arrangement mediating the exchange interaction.
Nevertheless, in CrI$_3$ the magnetic frustration due to third-neighbour interaction is rather weak~\cite{robust_fm_2015}, 
because of the honeycomb lattice adopted by the Cr cations, preventing the stabilization of a non-coplanar spin structure and favouring, instead, an easy-axis FM state. On the other hand, the presence of a strong two-site anisotropy 
 is necessary to obtain the topological triple-{\bf q} state when magnetic frustration is already strong enough to stabilize a non-collinear spin structure. In fact, in monolayer NiBr$_2$, 
for which we found a much weaker contribution of the two-site anisotropy to the exchange interaction, the ground-state is a single-{\bf q} helimagnetic state with no net scalar chirality (Fig. \ref{spin_cl_br}(a)). Its anisotropy still introduces frustration in the spins direction but it is energetically not strong enough to stabilize the skyrmionic pattern; 
thus, a complex non-coplanar helimagnetic texture takes place,
as detailed in Figs.~\ref{spin_cl_br}(a,c).
Furthermore, neither the single- nor the two-site anisotropies are strong enough to support a field-driven topological transition to a skyrmion lattice;  
rather, a single-{\bf q} conical cycloid state, shown in Fig.~\ref{spin_cl_br}(b,d), develops under applied field before a purely FM state is stabilized at $B/J_{1iso}\gtrsim 0.9$.\\

{\bf Discussion.}\\
In this work, we have identified the two-site anisotropy, arising from the short-range symmetric exchange interaction, as the driving mechanism for the stabilization of a spontaneous topological spin-structure. Specifically, we predict a spontaneous antibiskyrmion lattice below $T_c\simeq 30~K$ along with a field-induced topological transition in NiI$_2$ monolayer, representative of the class of 2D magnetic semiconductors.

We found that 
metastable multi-{\bf q} skyrmionic states, that may occur in frustrated magnets, can stabilize as ground-state with well-defined topology and chirality in the presence of competing two-site anisotropies characterized by non-coplanar principal axes. Such kind of additional frustration in the relative orientation of spins acts as an emergent chiral interaction, fixing the topology and the chirality of the localized spin textures and of the resulting skyrmion lattice, whereas its size and periodicity are mostly determined by competing isotropic exchanges. 
Interestingly, our findings are not limited to centrosymmetric systems, as the two-site anisotropy can be found in non-centrosymmetric magnets as well; its competition with the Dzyaloshinskii-Moriya interaction could thus also reveal interesting phenomena in many other systems, including the recently proposed Janus Cr(I,Br)$_3$ monolayer~\cite{laurent_cri3_2020}. 

In conclusion, the proposed mechanism, which we have predicted here in a Ni-halide monolayer, enlarges both the kind of magnetic interactions able to drive the stabilization 
of topological spin structures, and the class of materials able to host spontaneous skyrmionic lattices with definite chirality, 
including also magnetic semiconductors with short-range anisotropic interactions.

\vspace{1cm}

{\bf Methods}\\

\emph{First-principles calculations.}
Magnetic parameters of the interacting spin Hamiltonian~\ref{hamiltonian1} were calculated by performing first-principles simulations
within the Density Functional Theory (DFT), using the projector-augmented wave method as implemented in the VASP code~\cite{kresse_vasp, vasp_site}.
The following orbitals were considered as  valence states: Ni $3p$, $4s$ and $3d$, Cl $3s$ and $3p$, Br $4s$ and $4p$, and I $5s$ and $5p$.
The Perdew-Burke-Erzenhof (PBE) functional~\cite{pbe} within the generalized gradient approximation (GGA) was  employed to describe the exchange-correlation (xc) potential;
the plane wave cutoff energy was set to $600$~eV for NiCl$_2$ and NiBr$_2$, and $500$~eV for NiI$_2$, which is more then 130\% larger than the highest default value among the
involved elements.
The $U$ correction~\cite{dft+u} on the localized $3d$ orbitals of Ni atoms was  also included. Exchange energies reported in this article,
and used to run the MC simulations, were calculated by employing $U=1.8$~eV and $J=0.8$~eV within the Liechtenstein approach~\cite{liet_U}.
We also adopted the Dudarev approach~\cite{dudarev_U} to test the results solidity against different effective $U$ values; we choose $U$ equal to $1,2$ and $3$~eV and a fixed $J$
equal to $0$~eV. Results remain qualitatively the same: the $J_{3iso}/J_{1iso}$ and $J_{yz}/J_{1iso}$ ratios ({\em i.e.} the quantities with relevant physical meaning, even more than the absolute values of the interactions) are almost unaffected, as shown in Table SII in Supplementary.

We calculated the magnetic parameters reported in this paper
via the four-state energy mapping method, which is  explained in detail in
Refs.~\cite{mike_2013, laurent_cri3_2018,laurent_cri3_2020}, performing non-collinear DFT calculations plus spin-orbit coupling (SOC).
It is based on the use of large supercells, also allowing to exclude the coupling with unwanted distant neighbors.
By means of this method we can
obtain all the elements of the exchange tensor for a chosen magnetic pair,  thus gaining direct access to the symmetric anisotropic exchange part (the
two-site anisotropy) and the antisymmetric anisotropic part (the DM interaction) of the full exchange. 
In particular, we performed direct calculations on the magnetic Ni-Ni pair parallel to the $x$ direction, here denoted Ni$_0$-Ni$_1$ (Fig.~\ref{structures}a). The interaction between the five other  nearest-neighbor pairs can be evaluated via the three-fold rotational symmetry, as commented in Supplementary-Section SI.
In all our systems, the tensor turned out to be symmetric or, equivalently,
 excluding any anti-symmetric (DM-like) contribution. 
 
We performed calculations of the SIA, first- and second-neighbors interaction using a $5\times4\times1$ supercell; while a $6\times3\times1$ supercell for the estimate of the third- neighbors interaction. Such cells should exclude a significant influence from
next neighbors. We built supercells from the periodic repetition of the Ni$X_2$ monolayer unit cell, with lattice parameters and ionic positions optimized by performing
standard collinear DFT calculations with a ferromagnetic spin ordering. The obtained lattice parameters are: 3.49~\AA~ and 3.69~\AA~ for NiCl$_2$ and NiBr$_2$ respectively~\cite{danila},
which are in agreement (within 0.3\% uncertainty) with the values known for the bulk compounds, and 3.96~\AA~ for NiI$_2$, which is 1.5\% larger than the
3.89~\AA~ bulk value~\cite{mc_guire}. We thus checked the stability of the NiI$_2$ magnetic parameters by extracting them from a cell with lattice constant fixed to the
experimental value, not obtaining  significant changes (as reported in Table SIII in Supplementary). In all cases, the length of the out-of-plane axis, perpendicular to the monolayer plane, was fixed to 20.8~\AA, which
provides a distance of more than 17.5~\AA~ with respect to the periodic repetition of the layer along this direction.
The sampling of the Brillouin zone for the monolayer unit cell relied on a $18\times18\times1$ $k$-points mesh; meshes for the supercells have been chosen according the latter. \\

\emph{Monte Carlo simulations.} 
Monte Carlo (MC) calculations were performed using a standard Metropolis algorithm on $L\times L$ triangular lattices with periodic boundary conditions, with $L$ ranging between 10 and 60.
At each simulated temperature, we used 10$^5$ MC steps for thermalization and 5$\times$10$^5$ MC steps for statistical averaging. Average total energy, magnetization and specific heat have been calculated. We found that the lowest-energy configurations are obtained for supercells where $1/L$ is commensurate to the propagation vector $\bm q = 2 \cos^{-1}[(1+\sqrt{1-2J_{1iso}/J_{3iso}})/4]$ of the helical spin configuration induced by the $J_{3iso}/J_{1iso}$ magnetic frustration,~\cite{batista_2016}, i.e., $L=8$ for NiI$_2$ and $L=10$ for NiBr$_2$. Results are shown for calculations performed on 24$\times$24 and 30$\times$30 supercells for NiI$_2$ and NiBr$_2$, respectively, including  anisotropic first-neighbor and isotropic third-neighbor interactions reported in Table~\ref{exchange_j}.
Further insight on the magnetic configurations was obtained by evaluating the spin structure factor:
\begin{eqnarray}
S(\bm q) &=& \frac{1}{N}\,\sum_{\alpha=x,y,z}\,\left\langle \left\vert \sum_i S_{i,\alpha}\,e^{-i\bm q \cdot \bm r_i}\right\vert^2\right\rangle
\end{eqnarray}
where $\bm r_i$ denotes the position of spin $\bm S_i$ and $N=L^2$ is the total number of spins in the supercell used for MC simulations. The braket notation is used to denote the statistical average over the MC configurations. In order to assess the topological nature of the multiple-$q$ phase, we evaluate the topological charge (skyrmion number) of the lattice spin field of each supercell as $ \langle Q \rangle = \langle \sum_i \Omega_i\rangle$, where $\Omega_i$ is calculated for each triangular plaquette as~\cite{discrete_topological}:
\begin{eqnarray}
\tan\left(\frac{1}{2} \Omega_i \right)=\frac{\bm S_1\cdot \bm S_2 \times\bm S_3}{1+\bm S_1\cdot \bm S_2+\bm S_1\cdot \bm S_3+\bm S_2\cdot \bm S_3}
\end{eqnarray}
The corresponding topological susceptibility has been evaluated as:
\begin{eqnarray}
\chi_{Q} &=& \frac{\langle Q^2 \rangle -\langle Q\rangle^2}{k_B T}.
\end{eqnarray}
\\

{\bf Data Availability.}
Main results are reported in this article and related Supplementary Material. All other data that support the findings discussed in this study are available from the corresponding author upon reasonable request. \\

{\bf Acknowledgements}\\
This work was supported by the Nanoscience Foundries and Fine
Analysis (NFFA-MIUR Italy) project. P.B and S.P. acknowledge financial support from the Italian Ministry for Research and Education through PRIN-2017 projects
“Tuning and understanding Quantum phases in 2D materials - Quantum 2D” (IT-MIUR Grant No. 2017Z8TS5B) and “TWEET: Towards Ferroelectricity in two dimensions” (IT-MIUR Grant No. 2017YCTB59), respectively.
Calculations were performed on the high-performance computing (HPC) systems operated by CINECA,
supported by the ISCRA C (IsC66-I-2DFM, IsC72-2DFmF) and ISCRA B (IsB17-COMRED, grant HP10BSZ6LY) projects.
We thank Krisztian Palot\'as, Bertrand Dup\'e, Mario Cuoco, Hrishit Banerjee, Roser Valent\'i, Sang Wook Cheong, Sergey Artyukhin and Stefan Bl\"ugel for helpful and illuminating discussions. \\


\addcontentsline{toc}{section}{BIBLIOGRAPHY}
\begin{thebibliography}{100}
\bibitem{tokura_review}  Nagaosa, N. and Tokura, Y. Topological properties and dynamics of magnetic skyrmions. \emph{Nature Nanotechnology} 8, 899 (2013).
\bibitem{fert_review} Fert, A., Reyren, N., and Cros, V. Magnetic skyrmions: advances in physics and potential applications.
        \emph{Nat. Rev. Mater.} {\bf 2}, 17031 (2017).
\bibitem{parkin_racetrack} Parkin, S. and Yang S.-H. Memory on the racetrack \emph{Nature Nanotechnology} 10, 195 (2015)

\bibitem{robler_2006} R\"oBler, U. K., Bogdanov, A. N., and Pfleiderer, C. Spontaneous skyrmion ground states in magnetic metals. \emph{Nature} {\bf 442}, 797 (2006).
\bibitem{nagaosa_2009} Yi, S. D., Onoda, S., Nagaosa, N., and Han, J. H. Skyrmions and anomalous Hall effect in a Dzyaloshinskii-Moriya spiral magnet. 
\emph{Phys. Rev. B} {\bf 80}, 054416 (2009).
\bibitem{stefan_2011} Heinze, S., et al. Spontaneous atomic-scale magnetic skyrmion lattice in two dimensions. \emph{Nat. Phys.} {\bf 7}, 713 (2011).
\bibitem{tokura_report_2012} Seki, S., Yu, X. Z., Ishiwata, S., and Tokura, Y. Observation of Skyrmions in a Multiferroic Material. 
	\emph{Science} {\bf 336}, 198 (2012).
\bibitem{kristian_prb_2014} Simon E., Palot\'as, R\'ozsa, L., Udvardi, L., and Szunyogh, L. Formation of magnetic skyrmions with tunable properties in PdFe bilayer 
	deposited on Ir(111). \emph{Phys. Rev. B} {\bf 90}, 094410 (2014).
\bibitem{batista_2015} Lin, S.-Z., Saxena, A., and Batista, C. D. Skyrmion fractionalization and merons in chiral magnets with easy-plane anisotropy. 
	\emph{Phys. Rev. B} {\bf 91}, 224407 (2015).
%\bibitem{kristian_DW_2016} Vida, Gy. J., Simon, E., R\'ozsa, L., Palot\'as, K., and Szunyogh, L. Domain-wall profiles in Co/Ir$_n$/Pt(111) ultrathin films: 
%	Influence of the Dzyaloshinskii-Moriya interaction. \emph{Phys. Rev. B} {\bf 94}, 214422 (2016).
\bibitem{tokura_nanotech_2020} Peng, L., et al. Controlled transformation of skyrmions and antiskyrmions in a non-centrosymmetric magnet. 
	\emph{Nat. Nanotechnol.} (2020). https://doi.org/10.1038/s41565-019-0616-6.
\bibitem{okubo_prl} Okubo, T., Chung, S., and Kawamura, H. Multiple-q states and skyrmion lattice of the triangular-lattice Heisenberg 
	antiferromagnet under magnetic fields. \emph{Phys. Rev. Lett.} {\bf 108}, 017206 (2012).
%\bibitem{tokura_2012} X. Yu et al., Magnetic stripes and skyrmions with helicity reversals. \emph{Proc. Natl. Acad. Sci.} 109, 8856 (2012).
\bibitem{maxim_2015} Leonov, A. O. and Mostovoy, M. Multiply periodic states and isolated skyrmions in an anisotropic frustrated magnet. \emph{Nat. Commun.} {\bf 6}, 8275 (2015).
\bibitem{landau_2016} Lin, S.-Z. and Hayami, S. Ginzburg-Landau theory for skyrmions in inversion-symmetric magnets with competing interactions. 	\emph{Phys. Rev. B} {\bf 93}, 064430 (2016).
\bibitem{batista_2016} Hayami, S.,  Lin, S.-Z. and Batista, C. D. Bubble and skyrmion crystals in frustrated magnets with easy-axis anisotropy. \emph{Phys. Rev. B} {\bf 93}, 184413 (2016).
\bibitem{tokura_2019} Hirschberger, M. et al. Skyrmion phase and competing magnetic orders on
a breathing kagomé lattice. \emph{Nature Communications} {\bf 10}, 5831 (2019).
\bibitem{tokura_current_2014} Yu, X.Z., et al. Biskyrmion states and their current-driven motion in a layered manganite. \emph{Nat. Commun.} {\bf 5}, 3198 (2014). 
\bibitem{adv_mater_2Sk} W. Wang \emph{et al.} A Centrosymmetric Hexagonal Magnet with Superstable
Biskyrmion Magnetic Nanodomains in a Wide Temperature Range of 100–340 K. \emph{Adv. Mater} {\bf 28}, 6887 (2016).
\bibitem{current_induced_2017} Zhang, X. et al. Skyrmion dynamics in a frustrated ferromagnetic film and current-induced helicity 
locking-unlocking transition. 
	\emph{Nat. Commun.} {\bf 8}, 1717 (2017).
\bibitem{capic_2019} Capic, D., Garanin, D. A., and Chudnovsky, E. M. Biskyrmions Lattices in Centrosymmetric Magnetic Films. \emph{Phys. Rev. B} {\bf 100}, 014432 (2019).
\bibitem{ingrid_2019} G\"obel, B., Henk, J., and Mertig, I. Forming individual magnetic biskyrmions by merging two skyrmions in a centrosymmetric nanodisk. 
	\emph{Sci. Rep.} {\bf 9}, 9521 (2019).
\bibitem{motome_1} Ozawa, R., Hayami, S., and Motome Y. Zero-Field Skyrmions with a High Topological Number in Itinerant Magnets. \emph{Phys. Rev. Lett.} {\bf 118}, 147205 (2017).
\bibitem{motome_2} Hayami, S. and Motome, Y. Effect of magnetic anisotropy on skyrmions with high topological number in itinerant magnets. 
	\emph{Phys. Rev. B} {\bf 99}, 094420 (2019).
	%\bibitem{tokura_science_2019} Kurumaji et al., \emph{Science} 365, 914 (2019).
\bibitem{motome_3} Ozawa, R., Hayami, S., Barros, K., Chern, G.-W., Motome, Y. and Batista, C. D. Vortex Crystals with Chiral Stripes in Itinerant Magnets. \emph{J. Phys. Soc. Jpn.} {\bf 85}, 103703
(2016).
\bibitem{nibr2_1980} Adam, A., et al. Neutron diffraction study of the commensurate and incommensurate magnetic structures of niBr$_2$. 
	\emph{State Communications} {\bf 35}, 1 (1980). 
\bibitem{nii2_1981} Kuindersma, S.R., Sanchez, J. P., and Haas, C. Magnetic and structural investigations on NiI$_2$ and CoI$_2$. \emph{Physica} {\bf 111B}, 231 (1981).
\bibitem{nibr_tokura_2011} Tokunaga, Y., et al. Multiferroicity in NiBr$_2$ with long-wavelength cycloidal spin structure on a triangular lattice. 
	\emph{Phys. Rev. B} {\bf 84}, 060406(R) (2011).
\bibitem{nii2_tokura_2013} Kurumaji, T., et al. Magnetoelectric responses induced by domain rearrangement and spin structural change 
	in triangular-lattice helimagnets NiI$_2$ and CoI$_2$. \emph{Phys. Rev. B} {\bf 87}, 014429 (2013). 
\bibitem{mc_guire} McGuire, M. A. Crystal and Magnetic Structures in Layered, Transition Metal Dihalides and Trihalides. \emph{Crystals} {\bf 7}, 121 (2017).
\bibitem{kulish} Kulish, V. V. and Huang, W. Single-layer metal halides MX$_2$ (X = Cl, Br, I): stability and tunable magnetism from first principles and Monte Carlo simulations. \emph{J. Mater. Chem. C} {\bf 5}, 8734 (2017).
\bibitem{botana} Botana, A. S. and Norman M. R. Electronic structure and magnetism of transition metal dihalides: Bulk to monolayer. \emph{Phys. Rev. Materials} {\bf 3}, 044001 (2019).
\bibitem{babu} Babu, S., Prokes, S., Huang, Y. K., Radu, F., and Mishra S. K. Magnetic-field-induced incommensurate to collinear spin order transition in NiBr$_2$. \emph{J. Appl. Phys.} {\bf 125}, 093902 (2019). 
\bibitem{2D_hot} 2D magnetism gets hot. \emph{Nat. Nanotechnol.} {\bf 13}, 269 (2018).
\bibitem{CrSe} Zhang, Y., et al. Ultrathin Magnetic 2D Single‐Crystal CrSe. \emph{Adv. Mater} {\bf 31}, 1900056 (2019). 
\bibitem{VSe2} Wong, P. K. J. et al. Evidence of Spin Frustration in a Vanadium Diselenide Monolayer Magnet. \emph{Adv. Mater} {\bf 31}, 1901185 (2019).
\bibitem{adv_mater_2d} Li, H., Ruan, S., and Zeng,  Y.-J. Intrinsic Van Der Waals Magnetic Materials from Bulk to the 2D Limit: New Frontiers of Spintronics. \emph{Adv. Mater} {\bf 31}, 1900065 (2019).
\bibitem{helicity_tune_adv_mat} Hou, Z. et al. Current-Induced Helicity Reversal of a Single Skyrmionic Bubble Chain in a Nanostructured Frustrated Magnet. \emph{Adv. Mater} {\bf 32}, 1904815 (2019).
\bibitem{kristian_prb_2017} R\'ozsa, L. et al. Formation and stability of metastable skyrmionic spin structures with various topologies in an ultrathin film. \emph{Phys. Rev. B} 95, 094423 (2017).
\bibitem{bertrand_2016} Dup\'e, B., Kruse, C. N., Dornheim, T., and Heinze, S. How to reveal metastable skyrmionic spin structures by spin-polarized scanning tunneling microscopy. \emph{New J. Phys.} 18, 055015 (2016).
\bibitem{moriya1960} Moriya, T. Theory of magnetism of NiF$_2$. \emph{Phys. Rev. } {117}, 635 (1960).
\bibitem{mike_2013} Xiang, H., Lee, C., Koo, H-.J., Gong, X., and Whangbo, M-.H. Magnetic properties and energy-mapping analysis. \emph{Dalton Trans.} 42, 823 (2013).
\bibitem{laurent_cri3_2018} Xu, C., Feng, J., Xiang, H., and Bellaiche, L. Interplay between Kitaev interaction and single ion anisotropy in ferromagnetic CrI$_3$ and CrGeTe$_3$ monolayers. \emph{npj Comput. Mater.} 4, 57 (2018). 
\bibitem{robust_fm_2015} Zhang, W.-B., Qu, Q., Zhu, P., and Lam, C.-H. Robust intrinsic ferromagnetism and half semiconductivity in stable two-dimensional single-layer chromium trihalides. \emph{J. Mater. Chem. C} 3, 12457 (2015).
\bibitem{huang_2017} Huang, B. et al. Layer-dependent ferromagnetism in a van der Waals crystal down to the monolayer limit. \emph{Nature} 546, 270 (2017).
\bibitem{laurent_cri3_2020}Xu, C., et al. Topological spin texture in Janus monolayers of the chromium trihalides Cr(I,X)$_3$. \emph{Phys. Rev. B} {\bf 101}, 060404(R) (2020).
%\bibitem{merm_wag} Mermin, N. D. and Wagner, H. Absence of Ferromagnetism or Antiferromagnetism in One- or Two-Dimensional Isotropic Heisenberg Models. 
%	\emph{Phys. Rev. Lett.} {\bf 17}, 1133 (1966).
%\bibitem{kimura_spirals} Kimura, T. Spiral Magnets as magnetoelectrics. \emph{Annu. Rev. Mater. Res.} {\bf 37}, 387 (2007).
\bibitem{kresse_vasp} Kresse, G. and Furthm\"uller, J. Efficient iterative schemes for ab initio total-energy calculations using a plane-wave basis set.
        \emph{Phys. Rev. B} {\bf 54}, 11169 (1996).
\bibitem{vasp_site} VASP official website: http://cms.mpi.univie.ac.at/vasp
\bibitem{pbe} Perdew, J. P., Burke, K., and  Ernzerhof, M. Generalized Gradient Approximation Made Simple. \emph{Phys. Rev. Lett.} {\bf 77}, 3865 (1997).
\bibitem{dft+u} Rohrbach, A., Hafner, J., and Kresse, G. Electronic correlation effects in transition-metal sulfides.
        \emph{J. Phys.: Condens. Matter} {\bf 15}, 979 (2003).
\bibitem{liet_U} Liechtenstein, A. I. Anisimov, V. I., and Zaanen, J. Density-functional theory and strong interactions: Orbital ordering in Mott-Hubbard insulators.
        \emph{Phys. Rev. B} {\bf 52}, R5467(R) (1995).
\bibitem{dudarev_U} Dudarev, S. L., Botton, G. A., Savrasov, S. Y., Humphreys, C. J., and Sutton A. P.
        Electron-energy-loss spectra and the structural stability of nickel oxide: An LSDA+U study. \emph{Phys. Rev. B} {\bf 57}, 1505 (1998).
\bibitem{mike_2013} H. Xiang et al., \emph{Dalton Trans.} 42, 823 (2013).
\bibitem{danila} Amoroso, D. accepted in \emph{Nuovo Cimento C} {\bf 43}.
\bibitem{batista_2016} Hayami, S.,  Lin, S.-Z. and Batista, C. D. Bubble and skyrmion crystals in frustrated magnets with easy-axis anisotropy. \emph{Phys. Rev. B} {\bf 93}, 184413 (2016).
\bibitem{discrete_topological} Berg, B. and L\"uscher, M. Definition and statistical distributions of a topological number in the lattice O(3) $\sigma$-model. \emph{Nucl. Phys. B} {\bf 190}, 412-424 (1981)
\end{thebibliography}
\end{document}